# Flat-band-induced many-body interactions and exciton complexes in a layered semiconductor

Gabriele Pasquale[1,2,†], Zhe Sun[1,2,†,*], Kristians Cernevics[3], Raul Perea-Causin[5], Fedele Tagarelli[1,2], Kenji Watanabe[6], Takashi Taniguchi[7], Ermin Malic[4,5], Oleg V. Yazyev[3], Andras Kis[1,2,*]

[1]Institute of Electrical and Microengineering, École Polytechnique Fédérale de Lausanne (EPFL), CH-1015 Lausanne, Switzerland
[2]Institute of Materials Science and Engineering, École Polytechnique Fédérale de Lausanne (EPFL), CH-1015 Lausanne, Switzerland
[3]Institute of Physics, École Polytechnique Fédérale de Lausanne (EPFL), CH-1015 Lausanne, Switzerland
[4]Philipps-Universität Marburg, Department of Physics, Renthof 7, D-35032 Marburg, Germany
[5]Chalmers University of Technology, Department of Physics, 412 96 Gothenburg, Sweden
[6]Research Center for Functional Materials, National Institute for Materials Science, 1-1 Namiki, Tsukuba 305-0044, Japan
[7]International Center for Materials Nanoarchitectonics, National Institute for Materials Science, 1-1 Namiki, Tsukuba 305-0044, Japan

[†] These authors contributed equally to this work.
[*] Correspondence should be addressed to: Zhe Sun (zhe.sun@epfl.ch) and Andras Kis (andras.kis@epfl.ch)

**Interactions among a collection of particles generate many-body effects in solids resulting in striking modifications of material properties. The heavy carrier mass that yields strong interactions and gate control of carrier density over a wide range, make two-dimensional semiconductors an exciting playground to explore many-body physics. The family of III-VI metal monochalcogenides emerges as a new platform for this purpose due to its excellent optical properties and the flat valence band dispersion with a Mexican-hat-like inversion. In this work, we present a complete study of charge-tunable excitons in few-layer InSe by photoluminescence spectroscopy. From the optical spectra, we establish that free excitons in InSe are more likely to be captured by ionized donors due to the large exciton Bohr radius, leading to the formation of bound exciton complexes. Surprisingly, a pronounced redshift of the exciton energy accompanied by a decrease of the exciton**

**binding energy upon hole-doping reveals a significant band gap renormalization and dynamical screening induced by the presence of the Fermi reservoir. Our findings establish InSe as a reproducible and potentially manufacturable platform to explore electron correlation phenomena without the need for twist-angle engineering.**

**INTRODUCTION**

The behaviour of electrons in moiré superlattices has attracted tremendous interest as it constitutes a strongly correlated electron system where the mutual interaction energy exceeds the kinetic energy of carriers[1–5]. Two-dimensional (2D) layered materials are an ideal platform to realize such a system: the maximal kinetic energy of electrons in the moiré mini-bands is limited due to their small bandwidth while on the other hand, the reduced dielectric screening leads to strong Coulomb interactions among charge carriers. The crucial factor in further enhancing interactions is the large effective mass of carriers in the host materials, which is determined by the curvature of the relevant bands. III-VI metal monochalcogenides such as GaSe and InSe offer a valid alternative since the effective mass of holes could reach about $2m_0$ in the monolayer limit, where $m_0$ is the electron rest mass[6]. As a comparison, electrons and holes in monolayer transition metal dichalcogenides (TMDCs) have similar effective masses $m_e \approx m_h \approx 0.5m_0$[7]. The relative flat band dispersion of host materials also enables to form moiré superlattices in a wide range of twist angles[8].

Another intriguing property of III-VI metal monochalcogenides is the Mexican-hat-like inversion near the Γ point at the valence band edge. For example, as the γ-stacked InSe (Figure 1a) is thinned down to the monolayer limit, the valence band maximum (VBM) shifts from the Γ point, resulting in a van Hove singularity in the hole density of states (Figure S1). This kind of band dispersion only appears in the thin limit where the thickness of InSe is below seven layers[9]. Theory has predicted that tuning the Fermi level into the band inversion region could give rise to emergent phenomena such as half-metallicity and ferromagnetism[10–12]. With the growing scientific interest in flat-band materials like magic-angle graphene superlattices and twisted bilayer TMDCs, thin InSe emerges as a promising candidate for exploring strongly correlated systems without twist engineering. Although thin InSe is not rigorously a direct bandgap semiconductor, the modest momentum mismatch still allows the observation of thickness-dependent photoluminescence (PL) emissions from InSe excitons[13,14]. Combined with the excellent optical properties, we investigate the manifestations of many-body interactions through the PL measurements across a wide range of carrier densities.

## RESULTS AND DISCUSSION

### Charge-tunable excitons in few-layer InSe

Our device consists of a six-layer InSe (6L-InSe) encapsulated in hexagonal boron nitride (hBN), with a few-layer graphene bottom gate. A first-principles calculation of the band diagram, depicted in Figure 1b, shows that for 6L-InSe, the VBM is only about 5 meV higher than the Γ point in the valence band. Due to the relatively flat valence band dispersion, the effective mass of holes in InSe is much heavier than that of the electrons[6]. We calculate effective masses of electrons and holes of $m_e \approx 0.13\,m_0$ and $m_h \approx 0.9\,m_0$ for 6L-InSe (Supplementary Note 1). The corresponding reduced effective mass of excitons in 6L-InSe $m_r = \frac{m_e m_h}{m_e + m_h} \approx 0.11\,m_0$. The exciton binding energy $E_b$ and the Bohr radius $a_B$ are directly determined by it as $E_b = \frac{2 m_r e^4}{\epsilon_{eff}^2 \hbar^2}$ and $a_B = \frac{\epsilon_{eff} \hbar^2}{2 m_r e^2}$, where $e$ the elementary charge of an electron and $\epsilon_{eff}$ the effective dielectric constant. Excitons in few-layer InSe are therefore expected to possess a reduced binding energy $E_b$ and increased Bohr radius $a_B$ compared with excitons in TMDCs.

Figures 1c shows a schematic of our device A. We use a 720 nm pulsed laser with an 80 MHz repetition rate to excite the sample unless otherwise specified. All measurements were performed at 4.5 K. Figure 1e presents a typical PL spectrum recorded for gate voltage $V_g = 0$ V using a laser power $P = 50$ µW. The exciton peak at 1.48 eV has a linewidth of 15 meV, making it one of the narrowest recorded so far in few-layer InSe (Supplementary Note 3).

Up to now, excitons in thin InSe have been investigated without applying electrostatic gating[15–17]. Embedding a thin flake into a field-effect structure allows us to tune the Fermi level into the conduction or valence bands. Figure 2a shows a representative PL emission spectrum of our sample as a function of gating with multiple salient features. The bright and narrow peak located at 1.48 eV for $V_g$ between $-0.5$ V and 0 V corresponds to the exciton resonance $X_0$. As we increase the gate voltage above 0 V, the energy of the exciton peak is reduced by 7 meV ($X_-$) and a new broad peak ($D_-$) appears at 1.44 eV. On the other side, as we reduce the gate voltage to between $-5.8$ V and $-0.5$ V, the exciton peak evolves to 1.47 eV ($X_0'$) together with an intensity transfer to a broad peak centered at 1.35 eV ($D_0$). Further reduction in the gate voltage leads to a rapid decrease in the intensity of the 1.35 eV peak ($D_0$), and pronounced redshift of the 1.47 eV peak ($X_+$).

These features in the spectra are reproducible in other locations on the same sample (Supplementary Note 4) and a second device (device B). The latter device also has additional electrical contacts allowing us to determine the device charge configurations using electrical transport measurements (Supplementary Note 5). We find that for $V_g > 0$ V, we have electron doping (*n*-doped regime); for $-5.8 \text{ V} < V_g < 0$ V, the device is in the undoped regime and for $V_g < -5.8$ V, it is in the hole-doped (*p*-doped) regime. In accordance with the electrical measurements, the out-of-plane photocurrent ($I_g$ in Figure 1c), acquired during $V_g$-dependent PL measurements also exhibits three regimes depending on the gate voltage, further confirming the assignment of charge configurations (Supplementary Note 6). The upper panel of Figure 2b shows the PL count rate as a function of the gate voltage along the three vertical dashed lines shown in Figure 2a. When the $X_0$ emission is pronounced, both the $D_0$ and $D_-$ emissions become less intense. The lower panel of Figure 2b shows the peak energy as a function of the gate voltage. The $X_0$ emission for $-0.5\text{ V} < V_g < 0\text{ V}$ is characterized by a small blueshift (~ 7 meV). In the *p*-doped regime, the $X_+$ emission redshifts by 25 meV, as the carrier density is tuned to $1.3 \times 10^{12}\text{ cm}^{-2}$.

**Defect-bound excitons**

To understand the origins of the various spectral features, we first focus on the two peaks in the undoped regime. Figure 3a shows the emission intensity of $X_0'$ ($I_{X_0'}$) and $D_0$ ($I_{D_0}$) as a function of the excitation power for $V_g = -3\text{ V}$. Compared to the linear increase of $I_{X_0'}$ with excitation power, $I_{D_0}$ increases sublinearly and saturates at about $P_{sat} = 400\text{ μW}$. At high powers, $I_{X_0'}$ keeps growing continuously, and the low-energy tail of $X_0'$ leads to a further linear increase of $I_{D_0}$ when $P > P_{sat}$. We have also performed time-resolved PL (TRPL) measurements on the $X_0'$ and $D_0$ emissions separately (Supplementary Note 7). For the $D_0$ emission, a slow-decay component with a characteristic time of ~ 8 ns dominates. Based on the low saturation power as well as the long radiative lifetime, we identify the $D_0$ emission as originating from defect-bound excitons, which have been observed in other 2D semiconductors[18–21]. Additionally, our first-principles calculations suggest that selenium vacancies in InSe could induce localized states that are about 100-150 meV lower than the conduction band minimum (Figure S1), in agreement with the observed splitting between $X_0'$ and $D_0$ in PL spectra (~ 120 meV).

The presence of ionized defects could strongly influence the properties of optically generated quasi-particles, leading to the formation of bound exciton complexes (BXCs) which have been reported in bulk semiconducting crystals[22–26]. Within their radiative lifetime, free excitons have a probability to be affected by charged impurities, with a capture rate $\eta \propto n_D \sigma v_{th}$, where $n_D$ is the defect density, $\sigma$ is a capture cross-section proportional to $a_B^2$, and $v_{th}$ is the mean thermal velocity of the exciton[27]. $v_{th}$ satisfies the Maxwell-Boltzmann distribution, $v_{th} \propto \sqrt{k_B T}$, where $k_B$ is the Boltzmann constant and $T$ the temperature[28]. In experiments, the capture rate $\eta$ is proportional to the yield of the $D_0$ emission, which is defined as $\beta = I_{D_0}/(I_{D_0} + I_{X_0'})$. Figure 3b presents the temperature dependence of $\beta$ and the (normalized) total PL counts $I_{D_0} + I_{X_0'}$. At $T = 80$ K, the total PL intensity falls to 20 % due to the thermal dissociation of excitons into free electrons and holes. It allows us to estimate the exciton binding energy $E_b \sim k_B \cdot 80\,K = 7$ meV. As predicted, we observe that the yield of the $D_0$ emission $\beta$ increases at elevated temperatures. At $T = 50$ K, $\beta$ reaches 80%, suggesting that most of excitons are bound by defects.

**Exciton binding energy and absorption spectrum**

Emissions from defect-bound excitons allow us to obtain an absorption spectrum of free excitons and to extract the exciton binding energy. The excitonic absorption edge of InSe has been reported in bulk InSe[15,29]. However, for few-layer InSe, the exciton resonance cannot be observed by performing a reflectance measurement using a broadband white light source (Supplementary Note 8). Instead, we have carried out photoluminescence excitation (PLE) measurements by monitoring $I_{D_0}$ while tuning the excitation wavelength around the exciton resonance. Here, we use a narrow-linewidth tunable continuous-wave laser as a light source to excite the sample. Figure 3c shows the integrated PL intensity of the $D_0$ peak as a function of the excitation energy (blue shaded area) for $V_g = -3$ V and $-0.2$ V as well as the PL spectrum (orange shaded area) obtained at the same gate voltages. The PLE spectra are characterized by an absorption onset with an excitonic resonance at about 1.47 eV and are equivalent to the absorption spectrum of InSe excitons. Furthermore, when the laser energy is tuned below the exciton resonance but still far above the energy of $D_0$, $I_{D_0}$ almost vanishes. This indicates that the generation of defect-bound excitons requires the creation of free excitons in InSe and the subsequent relaxation from the exciton state to the defect-bound state (Figure 3d).

The presence of excitonic resonances influences the absorption edge of a semiconductor, which can be described by the Elliott model[30–32]. By applying the 2D Elliott model to the PLE data at $V_g = -3$ V and $-0.2$ V, we extract the exciton binding energy for $X_0'$ and $X_0$ to be 6 meV and 12 meV respectively (Supplementary Note 9), consistent with our temperature-dependent measurement presented in Figure 3b.

**Exciton complexes associated with donors**

Chalcogen vacancies in 2D semiconductors behave as electron donors due to unsaturated bonds[33–36]. Here, we consider how exciton properties change in the presence of neutral and ionized donors. The donor energy levels below the conduction band minimum can be filled or emptied depending on the Fermi level position. As shown in Figure 4a, when the Fermi level is tuned above the donor energy levels ($E_F > E_d$), all donor electrons remain bound to the donor ions since electrons tend to occupy all lower energy levels. Therefore, the donors are charge-neutral and this is referred to as freeze out (complete deionization)[37]. On the contrary, when the Fermi level is below the donor energy levels ($E_F < E_d$), donors will release the additional electrons to the reservoir (the contact) in the process of donor ionization[37].

The PLE experiments indicate that the optically excited electrons in the conduction band can relax and fill the donor levels. If the laser excitation is weak (exciton density $n_X < n_D$), a part of ionized donor sites will be temporarily occupied by these electrons up to a time scale that is on the order of the emission lifetime of $D_0$ (~ 8 ns), leading to dynamical deionization. If the laser excitation is sufficiently strong such that $n_X \gg n_D$, the donors are in the freeze-out state, even though $E_F < E_d$. Similarly, we call this process a dynamical freeze-out.

Figure 4b shows a table to illustrate different bound exciton complexes (BXCs) associated with neutral and ionized donors. When $E_d < E_F < E_c$, the InSe flake is electrostatically undoped and donors are in the freeze-out state, where $E_c$ ($E_v$) denotes the energy of the conduction (valence) band edge. The laser-induced electrons in the conduction band cannot relax to donor levels. We thereby assign the $X_0$ emission in $V_g \in [-0.5 \text{ V}, 0 \text{ V}]$ to the bare neutral exciton $|X\rangle$. When $E_v < E_F < E_d$, all donors should be ionized due to the Fermi level position. However, in the presence of laser excitation, the optically generated electrons can be captured by donor ions, forming a quasi-particle excited state $|De; h\rangle$ and resulting in the PL emission of $D_0$. Here, we use the notation $|De\rangle$ to express the neutral donor state and $|D\rangle$ to express the ionized donor state, where $D$ stands for the donor ion and $e/h$ stands for the electron/hole.

In the low-power regime, the free exciton density is lower than the donor density, and most of the donors are ionized. Electrons in free excitons experience additional attractive interaction with the ionized donors, inducing an energy redshift. Since this type of excitons is not composed of electrons that are captured by donors, we attribute the $X_0'$ emission in $V_g \in [-5.8 \text{ V}, -0.5 \text{ V}]$ to donor-dressed excitons $|D; X\rangle$. Both $|De; h\rangle$ and $|D; X\rangle$ return to the ionized donor state $|D\rangle$ after emitting photons. However, in the high-power regime, the number of excitons is much larger than that of donors, leading to a saturation of the $|De; h\rangle$ state. Moreover, the lifetime of $|De; h\rangle$ is much longer than the exciton lifetime, which means that compared to that of excitons, the dynamics of $|De; h\rangle$ can be regarded as quasi-static process. Because of dynamical freeze-out, the excitons are immersed in an electrostatic environment in which the donors play no role.

Based on the above discussion, we expect that the energy difference between $X_0$ and $X_0'$ ($\Delta E = E_{X_0} - E_{X_0'}$) should change with the number of ionized donors, and thus the excitation power. The red curve in Figure 4c depicts the energy difference $\Delta E$ as a function of the excitation power. The energy shift that vanishes at large powers ($P \gtrsim 800 \text{ μW}$) constitutes solid evidence for our interpretation. The carrier relaxations from the conduction band to the donor levels decrease the probability of generating bare excitons $|X\rangle$. As a result, the intensity of both $D_0$ and $X'_0$ are lower than $X_0$, as shown in Figure 2a. Finally, recalling that the binding energy of $X_0$ is double that of $X'_0$, as shown in Figure 3c, we verify that the presence of ionized donors can effectively reduce the exciton binding energy. A more detailed analysis can be found in Supplementary Note 10.

**Signatures of many-body interactions in p-doped regime**

Next, we focus on the prominent redshift in the *p*-doped regime. To begin with, we quantify the redshift energy $E_{red}$ as a function of the Fermi energy $E_F$ (Supplementary Note 11). We find the change of $E_{red}$ is about one order of magnitude larger than the change of the Fermi energy $dE_{red}/dE_F \approx -10$ (Figure 5a). Here, the minus sign denotes the redshift of the PL energy.

Generally, the energy of PL emission $E_{PL}$ can be described as $E_{PL} = E_g - E_b$, where $E_g$ is the single-particle band gap. In the presence of a Fermi reservoir, dynamical screening of the electron-hole interactions leads to a lower binding energy $E_b$, as the system approaches the Mott transition[38]. As a result, we expect that this gives rise to a vanishing PL intensity together with a blueshift as the carrier density is increased. For excitons in InSe, we expect a modest blueshift since $E_b$ is a small quantity.

On the other hand, many-body screening also renormalizes the particle self-energies and results in a reduced $E_g$, which is referred to as the band gap renormalization (BGR)[39,40]. The overall decrease of $E_g$ can be represented in the following form: $\delta E_g = \Sigma_{SX} + 2\Sigma_{CH}$, where $\Sigma_{SX}$ is the screened exchange term and $\Sigma_{CH}$ is the Coulomb-hole term[41]. The former is a consequence of the reduced Coulomb repulsion arising from the exchange interaction due to the Pauli exclusion principle. The latter describes the energy decrease caused by the depletion shell around a charge carrier. The screened exchange term is on the order of the Fermi energy $\Sigma_{SX} \approx -1/2\, E_F$ [39], and hence cannot explain the large redshift in the *p*-doped regime.

The Coulomb-hole term is sensitive to the effective mass of particles in a Fermi liquid. The effect of many-body screening can be represented by introducing a dynamical dielectric function $\epsilon(q, \omega) = V_q/W_q$, where $q$ is the wavenumber, $V_q$ is the bare Coulomb potential and $W_q$ is the screened Coulomb potential in momentum space. This allows us to calculate the Coulomb-hole term via $2\Sigma_{CH} = \sum_q (W_q - V_q)$. For a 2D system in the static limit, $\epsilon(q, 0) = 1 + \kappa/q$. $\kappa$ is the screening wavenumber which takes the form of $\kappa = (e^2/2\epsilon_0\epsilon_r)(\partial n/\partial \mu)$, where $\mu$ is the chemical potential, $\partial n/\partial \mu$ is the DOS, $\epsilon_0$ is the vacuum permittivity and $\epsilon_r$ is the dielectric constant of InSe. It is straightforward to notice that a heavier carrier mass (larger DOS) leads to a larger screening wavenumber and a stronger BGR. It explains our observation that the PL spectrum shows a strong redshift upon hole-doping, whereas it almost remains constant upon electron-doping (Figure S4f).

According to this picture, we calculate the Coulomb-hole term as a function of the Fermi energy for two carrier masses $m = 0.1\ m_0$ and $m = m_0$. From the calculation results depicted in Figure 5b, we expect that the BGR induced by a hole reservoir ($m = m_0$, blue curve) is about 4 - 5 times larger than that induced by an electron reservoir ($m = 0.1\ m_0$, red curve). Since the calculations are performed in the static approximation, the Coulomb-hole term is overestimated[39] (Supplementary Note 11). Nevertheless, the BGR model can estimate the redshift within a correct order of magnitude. The value of $\Sigma_{CH}$ shows a saturation when $E_F > 1$ meV. However, our experiment results do not exhibit a clear saturation for the Fermi energies up to $E_F \sim 3$ meV. This can be explained by the fact that our InSe flake is thicker than a monolayer. The inverse of the screening wavenumber $\kappa$ (Figure S10), which can be regarded as the screening length in 2D, is on the order of 0.3 nm, similar to the thickness of a monolayer. This means that for thicker flakes (~5 nm for our sample), the 2D screening model is no longer precise. We develop the 2D model into a 3D model, by replacing the expression of the dynamical dielectric function and screening wavenumber with $\epsilon(q, 0) = 1 + \kappa^2/q^2$ and $\kappa = $ $\sqrt{(e^2/\epsilon_0\epsilon_r)(\partial n/\partial \mu)}$ (Supplementary Note 11). The green curve in Figure 5c presents the calculation result using the 3D model that predicts a more gradual change of $\Sigma_{CH}$ with increasing Fermi energy.

**Outlook**

Our results and analysis elucidate the origins of exciton species in different charge regimes and provide an insight into the optical signatures of strong many-body interactions between charge carriers. So far, the research on bound exciton complexes has been conducted mainly on bulk semiconductors. 2D materials have recently emerged as a highly-tunable system that allows modifying the carrier density over a large range. Hence, our findings motivate further exploration of bound exciton complexes in charge-tunable 2D semiconductors. Observing intense emission from defect-bound excitons motivates the search for individual InSe-hosted single-photon emitters, similar to those that have been discovered in other layered semiconductors[42,43]. Another future research direction would be to explore optical signatures of strongly correlated phenomena by tuning the Fermi level into the band inversion region of a thinner InSe flake.

**MATERIALS AND METHODS**

**Sample fabrication**

The heterostructures presented in this work are fabricated by a three-step dry transfer technique. First, the hBN and few-layer graphene (NGS) building blocks are exfoliated on silicon substrates with a 280 nm oxide layer to optimize the optical contrast. The bottom hBN flake is picked up with the dry-transfer method and released on top of the few-layer graphene (FLG) bottom gate. The few-layer InSe (HQ Graphene) flakes are exfoliated on PDMS (Gelpak), identified by optical contrast, and transferred on top of the previously fabricated hBN/FLG stack. Top hBN and FLG contacts are subsequently picked up, carefully aligned, and released on top of the InSe flake, fully encapsulating the heterostructure. All these steps are performed in argon environment to avoid any contamination. Once completed, the sample is annealed at 340 ºC in high vacuum at $10^{-6}$ mbar for 6h. Finally, electrical contacts are fabricated by e-beam lithography and metal evaporation (2/100 nm Ti/Au).

**Optical and electrical measurements**

All measurements shown in this work were carried out under vacuum at 4.5 K unless specified otherwise. PL measurements were performed by focusing a laser on a spot of about 1 µm diameter on the sample. Multiple laser sources have been used for this purpose: a tunable femtosecond pulsed laser (Coherent Chameleon) and a narrow-linewidth tunable continues wave laser (MSquared), which is used for PLE measurements (Figure 3c). The incident power was varied from 1 µW to 5 mW for power dependence measurements (Figure 3a) and kept at 50 µW for the PL measurements shown in Figure 2. Transport measurements were carried out at room temperature with the Keithley 2636 Sourcemeter. TRPL measurements were performed by sending the collected photons to an APD (Excelitas Technologies, SPCM-AQRH-16). A time-correlated photon counting module (TCPCM) with a resolution of 12 ps r.m.s. (PicoQuant, PicoHarp 300) is connected to the output of the APD, to measure the arrival time of photons. The instrument resolution (full width at half maximum) is 300 ps.

**First-principles calculations**

Our first-principles calculations were performed at the density-functional theory level as implemented in VASP[44]. We used the semilocal PBE functional[45] for all structure relaxations and the hybrid HSE06[46] for accurate band gap estimations of pristine systems. To obtain the correct band gap for defect calculations containing large supercells (where hybrid functional couldn't be applied) we used a modified Becke-Johnson exchange potential in combination with LDA-correlation[47,48]. Electron-core interactions were described through the projector augmented wave (PAW) method[49,50], while Kohn-Sham wave functions were expanded in a plane-wave basis set with a cutoff on the kinetic energy of 400 eV. All structures were subjected to periodic boundary conditions and a 4×4 supercell geometry was applied for the defect calculations. Vacuum layer of 10 Å perpendicular to the layers was used to prevent interaction between replica images. The integration over the Brillouin zone was carried out using a 10×10×1 and 3×3×1 k-point mesh for pristine and defective systems respectively. Atomic positions and lattice constants were optimized using the conjugate gradient method, where the total energy and atomic forces were minimized. The convergence criterion for energy was chosen to be $10^{-5}$ eV and the maximum force acting on each atom was less than 0.01 eV/Å relaxation.

**FIGURES**

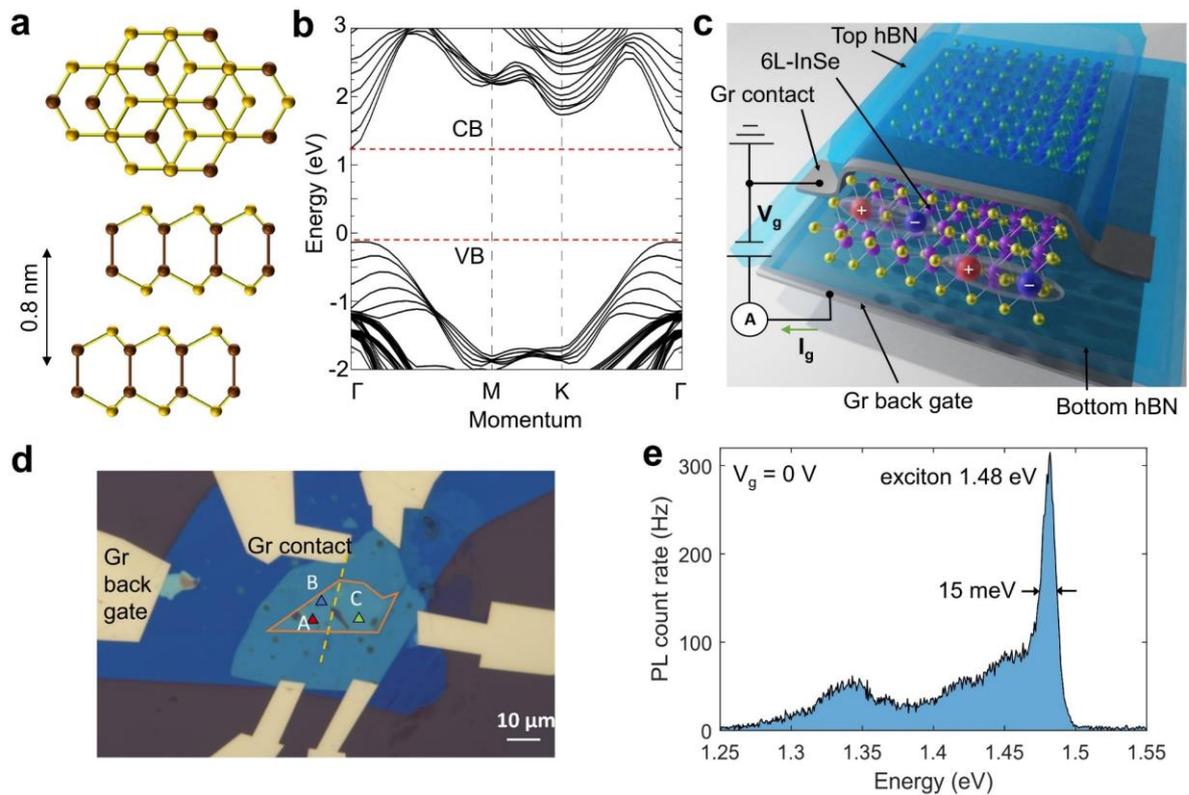

**Figure 1. Few-layer InSe in a field-effect device structure.** (**a**) Top and side views of γ-stacked bilayer InSe. Brown spheres: Indium atom; yellow spheres: Selenium atom. The interlayer distance is about 0.8 nm. (**b**) Calculated band diagram for 6L-InSe. CB: conduction band; VB: valence band. (**c**) Schematic of the hBN-encapsulated 6L-InSe device (device A) with a few-layer graphene contact and bottom gate. (**d**) Optical microscope image of device A. The orange contour encloses the region of the InSe flake. The dashed line indicates the position of the few-layer graphene contact on the flake. Scale bar, 10 μm. (**e**) PL count rate as a function of the emission energy at $V_g = 0$ V using $P = 50$ μW, measured at position A in Fig. 1(d).

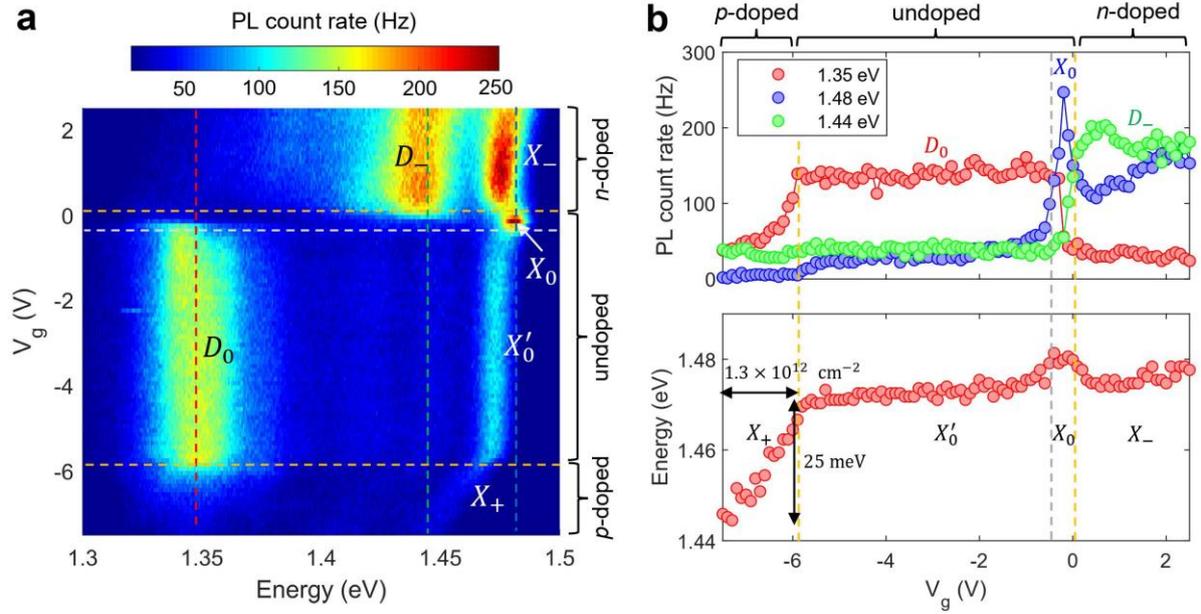

**Figure 2. Charge-tunable excitons in few-layer InSe.** (a) PL count rate as a function of the emission energy and gate voltage using $P = 50$ μW, measured at position A in Fig. 1(d). The horizontal yellow dashed lines indicate different charging regimes. (b) Upper panel: PL count rate as a function of the gate voltage along the three vertical dashed lines in Fig. 2(a). Red: $D_0$; blue: $X_0$; green: $D_-$. Lower panel: peak energy as a function of the gate voltage. The vertical yellow dashed lines indicate different charging regimes. Here, 'undoped' means un-intentionally doped.

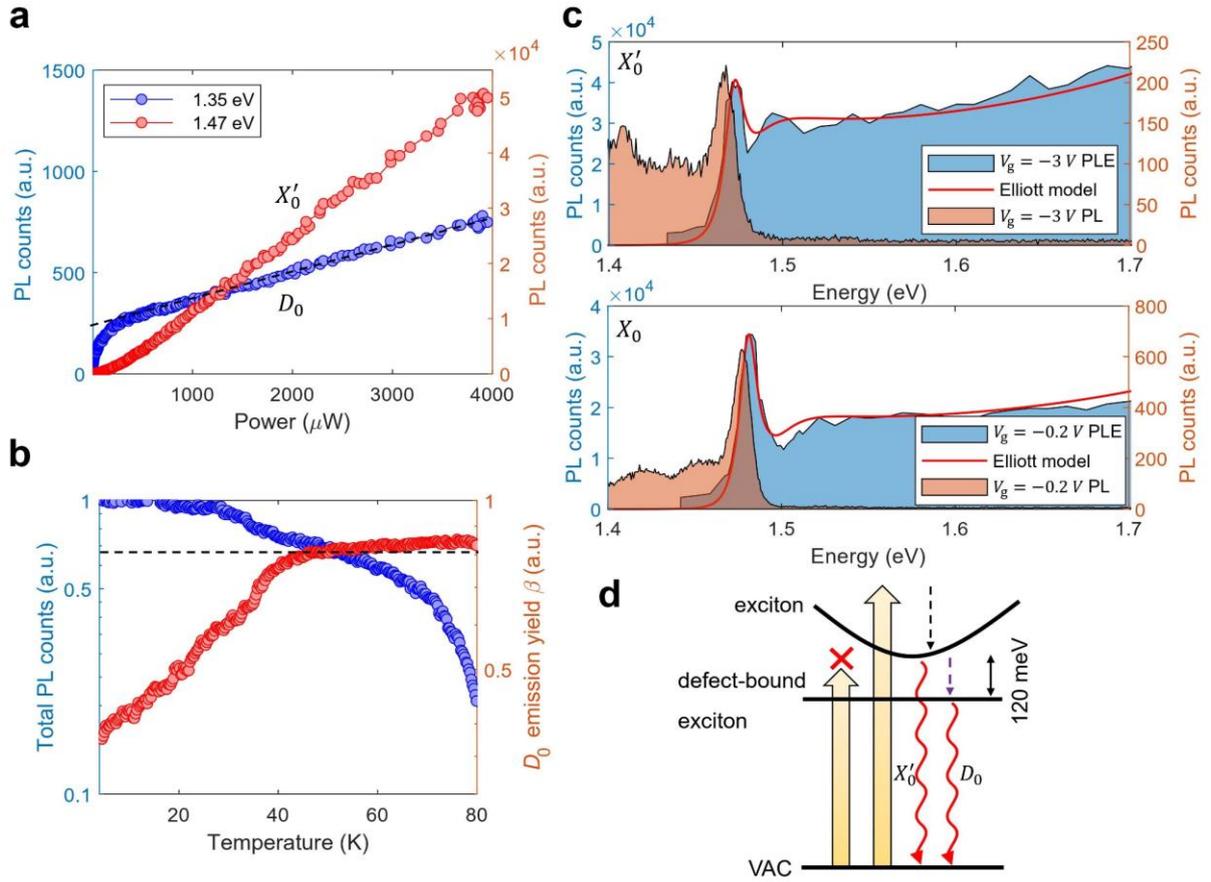

**Figure 3. Defect-bound exciton and photoluminescence excitation spectrum. (a)** Emission intensity of $X_0'$ (blue) and $D_0$ (red) peaks as a function of the laser power. The black dashed line indicates the linear background originating from the low-energy tail of $X_0'$ at high power. **(b)** Blue: (normalized) total PL counts of $X_0'$ and $D_0$ as a function of temperature using $P = 50$ μW. Red: PL emission yield of $D_0$, defined as $\beta = I_{D_0}/(I_{D_0} + I_{X_0'})$, as a function of temperature. **(c)** Integrated PL intensity of $D_0$ as a function of excitation energy (PLE spectrum, blue shaded area) and PL intensity as a function of emission energy (orange shaded area) at $V_g = -3$ V and $-0.2$ V. The red solid lines are fits using the 2D Elliott model. For $X_0$ ($V_g = -0.2$ V), the fitting parameters are $E_g = 1.493$ eV and $E_b = 12$ meV; For $X_0'$ ($V_g = -3$ V), the fitting parameters are $E_g = 1.478$ eV and $E_b = 6$ meV. **(d)** Schematic of the exciton state and defect-bound exciton state in the quasi-particle picture. Yellow arrows: laser excitation; dashed arrows: relaxations; wave arrows: radiative emissions. VAC stands for the vacuum state.

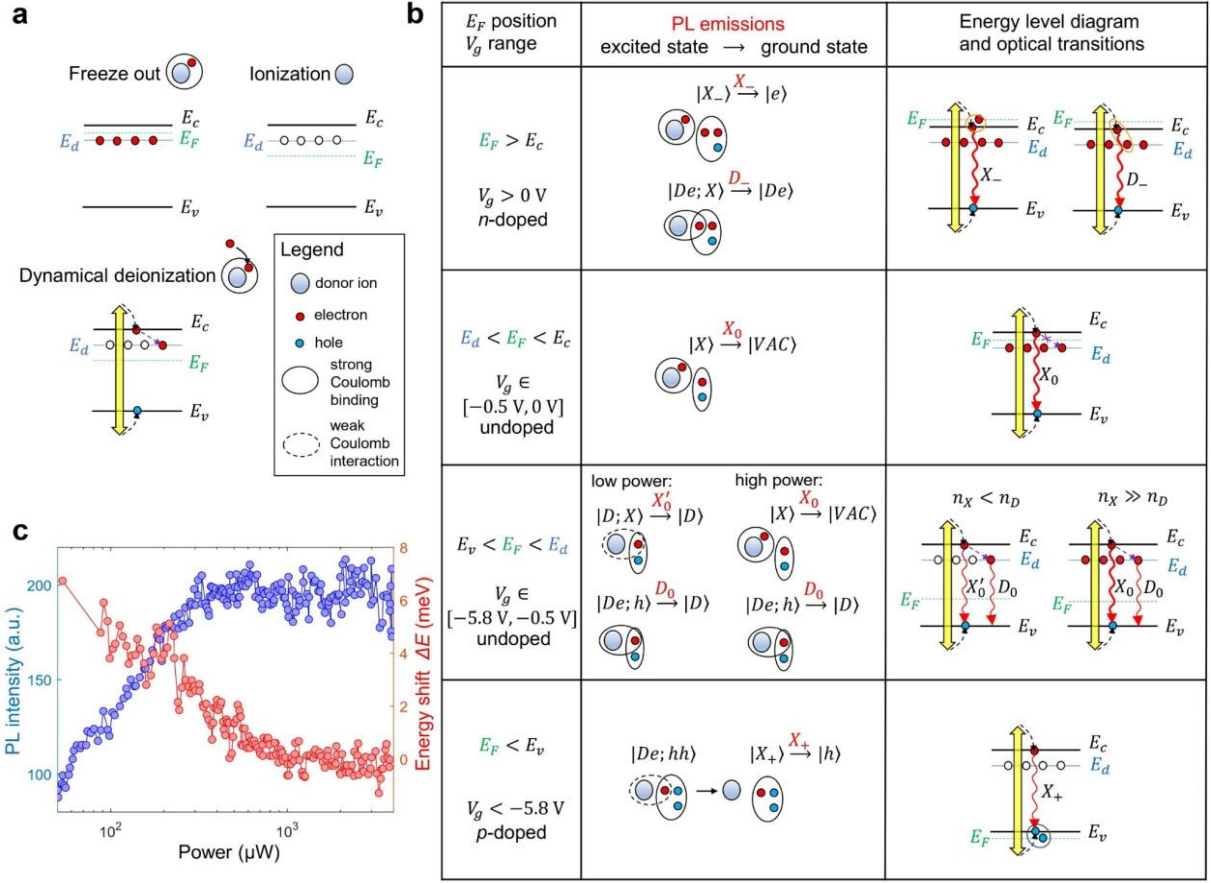

**Figure 4. Bound exciton complexes associated with neutral and ionized donors. (a)**. Schematic of donor freeze-out, ionization and dynamical deionization. Yellow arrow: laser excitation; dashed arrow: carrier relaxations. **(b).** Table illustrating different quasi-particle (excited and ground) states, relevant PL emissions and optical transitions for different Fermi level positions. The notation $|...\rangle$ indicates the quasi-particle state. $D$: donor ion; $e$: electron; $h$: hole; $X$: exciton; $VAC$: vacuum state. The red words above $\rightarrow$ indicate the names for PL emissions defined in Fig. 2a. **(c).** Red: energy shift between $X_0$ and $X_0'$, $\Delta E = E_{X_0} - E_{X_0'}$ as a function of the laser power. Blue: PL intensity of defect-bound excitons ($D_0$) as a function of the laser power at $V_g = -3$ V, after removing the linear background.

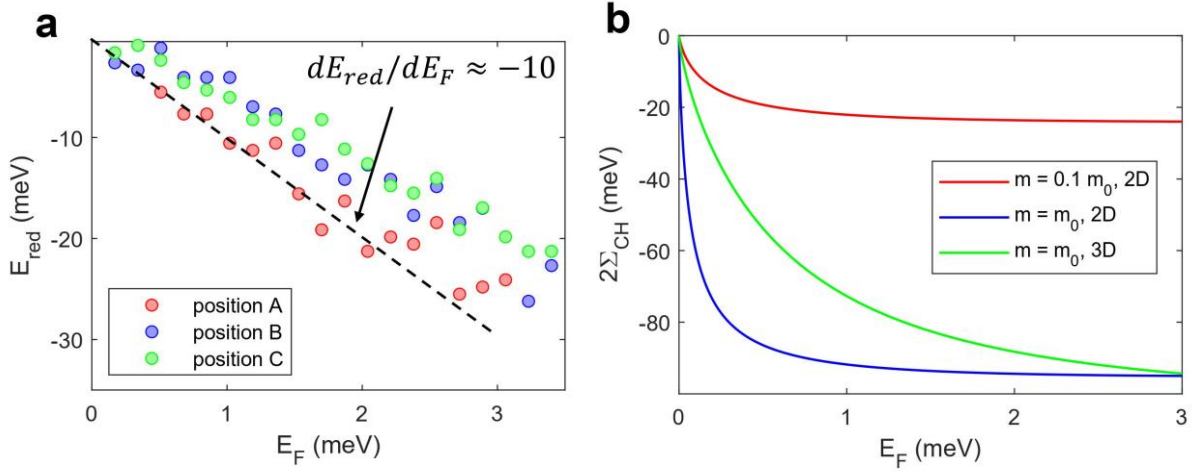

**Figure 5. Redshift in the *p*-doped regime and band gap renormalization. (a)** Redshift energy $E_{red}$ in the *p*-doped regime as a function of the Fermi energy $E_F$, measured on positions A, B & C in Fig. 1(d). **(b)** Calculated Coulomb-hole term $2\Sigma_{CH}$ as a function of the Fermi energy $E_F$ for $m = 0.1\, m_0$ and $m = m_0$ in 2D (red and blue curve), and for $m = m_0$ in 3D (green curve).

**REFERENCES**

1. Cao, Y. *et al.* Unconventional superconductivity in magic-angle graphene superlattices. *Nature* **556**, 43–50 (2018).
2. Cao, Y. *et al.* Correlated insulator behaviour at half-filling in magic-angle graphene superlattices. *Nature* **556**, 80–84 (2018).
3. Tang, Y. *et al.* Simulation of Hubbard model physics in $WSe_2/WS_2$ moiré superlattices. *Nature* **579**, 353–358 (2020).
4. Regan, E. C. *et al.* Mott and generalized Wigner crystal states in $WSe_2/WS_2$ moiré superlattices. *Nature* **579**, 359–363 (2020).
5. Wang, L. *et al.* Correlated electronic phases in twisted bilayer transition metal dichalcogenides. *Nat. Mater.* **19**, 861–866 (2020).
6. Ben Aziza, Z. *et al.* Valence band inversion and spin-orbit effects in the electronic structure of monolayer GaSe. *Phys. Rev. B* **98**, 115405 (2018).
7. Kormányos, A. *et al.* k·p theory for two-dimensional transition metal dichalcogenide semiconductors. *2D Mater.* **2**, 022001 (2015).
8. Tao, S. *et al.* Designing Ultra-Flat Bands in Twisted Bilayer Materials at Large Twist Angles without specific degree. *J. Am. Chem. Soc.* **144**, 3949–3956 (2022).
9. Magorrian, S. J., Zólyomi, V. & Fal'ko, V. I. Electronic and optical properties of two-dimensional InSe from a DFT-parametrized tight-binding model. *Phys. Rev. B* **94**, 245431 (2016).
10. Cao, T., Li, Z. & Louie, S. G. Tunable Magnetism and Half-Metallicity in Hole-Doped Monolayer GaSe. *Phys. Rev. Lett.* **114**, 236602 (2015).
11. Feng, W., Guo, G.-Y. & Yao, Y. Tunable magneto-optical effects in hole-doped group-IIIA metal-monochalcogenide monolayers. *2D Mater.* **4**, 015017 (2016).
12. Stepanov, E. A. *et al.* Coexisting charge density wave and ferromagnetic instabilities in monolayer InSe. *ArXiv210701132 Cond-Mat* (2021).
13. Brotons-Gisbert, M. *et al.* Nanotexturing To Enhance Photoluminescent Response of Atomically Thin Indium Selenide with Highly Tunable Band Gap. *Nano Lett.* **16**, 3221–3229 (2016).
14. Bandurin, D. A. *et al.* High electron mobility, quantum Hall effect and anomalous optical response in atomically thin InSe. *Nat. Nanotechnol.* **12**, 223–227 (2017).
15. Shubina, T. V. *et al.* InSe as a case between 3D and 2D layered crystals for excitons. *Nat. Commun.* **10**, 3479 (2019).
16. Zultak, J. *et al.* Ultra-thin van der Waals crystals as semiconductor quantum wells. *Nat. Commun.* **11**, 125 (2020).
17. Venanzi, T. *et al.* Photoluminescence dynamics in few-layer InSe. *Phys. Rev. Mater.* **4**, 044001 (2020).
18. Moody, G. *et al.* Microsecond Valley Lifetime of Defect-Bound Excitons in Monolayer $WSe_2$. *Phys. Rev. Lett.* **121**, 057403 (2018).
19. Shang, J. *et al.* Revealing electronic nature of broad bound exciton bands in two-dimensional semiconducting $WS_2$ and $MoS_2$. *Phys. Rev. Mater.* **1**, 074001 (2017).
20. Rivera, P. *et al.* Intrinsic donor-bound excitons in ultraclean monolayer semiconductors. *Nat. Commun.* **12**, 871 (2021).
21. He, M. *et al.* Valley phonons and exciton complexes in a monolayer semiconductor. *Nat. Commun.* **11**, 618 (2020).
22. Petelenz, P. & Smith Jr., V. H. Binding energy of the Wannier exciton – ionized donor complex in the CdS crystal. *Can. J. Phys.* **57**, 2126–2131 (1979).
23. Merz, J. L., Kukimoto, H., Nassau, K. & Shiever, J. W. Optical Properties of Substitutional Donors in ZnSe. *Phys. Rev. B* **6**, 545–556 (1972).

24. Šantic, B. *et al.* Ionized donor bound excitons in GaN. *Appl. Phys. Lett.* **71**, 1837–1839 (1997).
25. Meyer, B. K., Sann, J., Lautenschläger, S., Wagner, M. R. & Hoffmann, A. Ionized and neutral donor-bound excitons in ZnO. *Phys. Rev. B* **76**, 184120 (2007).
26. Bogardus, E. H. & Bebb, H. B. Bound-Exciton, Free-Exciton, Band-Acceptor, Donor-Acceptor, and Auger Recombination in GaAs. *Phys. Rev.* **176**, 993–1002 (1968).
27. Peyghambarian, N., Koch, S. W. & Mysyrowicz, A. *Introduction to Semiconductor Optics*. (Prentice Hall, 1993).
28. Harris, C. *et al.* Temperature dependence of exciton-capture at impurities in GaAs/Al$_x$Ga$_{(1-x)}$As quantum wells. *J. Phys. IV Proc.* **03**, C5-171-C5-174 (1993).
29. Camassel, J., Merle, P., Mathieu, H. & Chevy, A. Excitonic absorption edge of indium selenide. *Phys. Rev. B* **17**, 4718–4725 (1978).
30. Elliott, R. J. Intensity of Optical Absorption by Excitons. *Phys. Rev.* **108**, 1384–1389 (1957).
31. Passarelli, J. V. *et al.* Tunable exciton binding energy in 2D hybrid layered perovskites through donor–acceptor interactions within the organic layer. *Nat. Chem.* **12**, 672–682 (2020).
32. Neutzner, S. *et al.* Exciton-polaron spectral structures in two-dimensional hybrid lead-halide perovskites. *Phys. Rev. Mater.* **2**, 064605 (2018).
33. Cho, K. *et al.* Electrical and Optical Characterization of MoS$_2$ with Sulfur Vacancy Passivation by Treatment with Alkanethiol Molecules. *ACS Nano* **9**, 8044–8053 (2015).
34. Shen, P.-C. *et al.* Healing of donor defect states in monolayer molybdenum disulfide using oxygen-incorporated chemical vapour deposition. *Nat. Electron.* **5**, 28–36 (2022).
35. Zhou, W. *et al.* Intrinsic Structural Defects in Monolayer Molybdenum Disulfide. *Nano Lett.* **13**, 2615–2622 (2013).
36. Zhang, S. *et al.* Defect Structure of Localized Excitons in a WSe$_2$ Monolayer. *Phys. Rev. Lett.* **119**, 046101 (2017).
37. Neamen. *Semiconductor Physics And Devices*. (McGraw-Hill Education (India) Pvt Limited, 2003).
38. Wang, Z., Zhao, L., Mak, K. F. & Shan, J. Probing the Spin-Polarized Electronic Band Structure in Monolayer Transition Metal Dichalcogenides by Optical Spectroscopy. *Nano Lett.* **17**, 740–746 (2017).
39. Scharf, B., Tuan, D. V., Žutić, I. & Dery, H. Dynamical screening in monolayer transition-metal dichalcogenides and its manifestations in the exciton spectrum. *J. Phys.: Condens. Matter* **31**, 203001 (2019).
40. Van Tuan, D. *et al.* Probing many-body interactions in monolayer transition-metal dichalcogenides. *Phys. Rev. B* **99**, 085301 (2019).
41. Haug, H. & Koch, S. W. *Quantum Theory of the Optical and Electronic Properties of Semiconductors*. (World Scientific, 2009).
42. Srivastava, A. *et al.* Optically active quantum dots in monolayer WSe$_2$. *Nat. Nanotechnol.* **10**, 491–496 (2015).
43. Koperski, M. *et al.* Single photon emitters in exfoliated WSe$_2$ structures. *Nat. Nanotechnol.* **10**, 503–506 (2015).
44. Kresse, G. & Furthmüller, J. Efficient iterative schemes for ab initio total-energy calculations using a plane-wave basis set. *Phys. Rev. B* **54**, 11169–11186 (1996).
45. Perdew, J. P., Burke, K. & Ernzerhof, M. Generalized Gradient Approximation Made Simple. *Phys. Rev. Lett.* **77**, 3865–3868 (1996).
46. Heyd, J., Scuseria, G. E. & Ernzerhof, M. Hybrid functionals based on a screened Coulomb potential. *J. Chem. Phys.* **118**, 8207–8215 (2003).

47. Becke, A. D. & Johnson, E. R. A simple effective potential for exchange. *J. Chem. Phys.* **124**, 221101 (2006).
48. Tran, F. & Blaha, P. Accurate Band Gaps of Semiconductors and Insulators with a Semilocal Exchange-Correlation Potential. *Phys. Rev. Lett.* **102**, 226401 (2009).
49. Kresse, G. & Joubert, D. From ultrasoft pseudopotentials to the projector augmented-wave method. *Phys. Rev. B* **59**, 1758–1775 (1999).
50. Blöchl, P. E. Projector augmented-wave method. *Phys. Rev. B* **50**, 17953–17979 (1994).

**ACKNOWLEDGEMENTS**

We acknowledge many helpful discussions with Dr. Aymeric Delteil, Juan Francisco Gonzalez Marin and Edoardo Lopriore. We acknowledge the support in microfabrication and e-beam lithography from EPFL Centre of MicroNanotechnology (CMI) and thank Z. Benes (CMI) for help with electron-beam lithography. **Fundings:** We acknowledge support from Swiss National Science Foundation (grant nos. 175822, 177007 and 164015) the European Union's Horizon 2020 research and innovation program under grant agreements 785219 and 881603 (Graphene Flagship Core 2 and Core 3). K.W. and T.T. acknowledge support from the Elemental Strategy Initiative conducted by the MEXT, Japan (Grant Number JPMXP0112101001) and JSPS KAKENHI (Grant Numbers JP19H05790 and JP20H00354). E. M. acknowledges funding from Deutsche Forschungsgemeinschaft via CRC 1083 and the Europe Unions Horizon 2020 research and innovation program under grant agreement No. 881603 (Graphene Flagship). **Author contributions:** A.K. conceived and supervised the project. Z.S. and G.P. performed the optical and electrical measurements. G.P. fabricated the samples. Z.S. built the experimental setups. Z.S. analyzed and explained the data with input from all authors. K.C. performed first-principles calculations with input from O.Y. K.W. and T.T. grew the h-BN crystals. Z.S. and A.K. wrote the manuscript with input from all authors. Competing interests: The authors declare no competing interests. **Data and materials availability:** All data needed to evaluate the conclusions in the paper are present in the paper and/or the Supplementary Materials. The data that support the plots within this paper and other findings of this study are available upon reasonable request.